# Dispersion Engineered Metasurfaces for Broadband, High-NA, High-Efficiency, Dual-Polarization Analog Image Processing

Michele Cotrufo[1,2,†,*], Akshaj Arora[1,3,†], Sahitya Singh[1,3] and Andrea Alù[1,3,*]

[1]*Photonics Initiative, Advanced Science Research Center, City University of New York, New York, NY 10031, USA*

[2]*The Institute of Optics, University of Rochester, Rochester, NY 14627, USA*

[3]*Physics Program, Graduate Center of the City University of New York, New York, NY 10016, USA*

[†]*These authors contributed equally*

*\* aalu@gc.cuny.edu, \* mcotrufo@optics.rochester.edu*

## Abstract

*Optical metasurfaces performing analog image processing – such as spatial differentiation and edge detection – hold the potential to reduce processing times and power consumption, while avoiding bulky 4F lens systems. However, current designs have been suffering from trade-offs between spatial resolution, throughput, polarization asymmetry, operational bandwidth, and isotropy. Here, we show that dispersion engineering provides an elegant way to design metasurfaces where all these critical metrics are simultaneously optimized. We experimentally demonstrate silicon metasurfaces performing isotropic and dual-polarization edge detection, with numerical apertures above 0.35 and spectral bandwidths of 35 nm around 1500 nm. Moreover, we introduce quantitative metrics to assess the efficiency of these devices. Thanks to the low loss nature and dual-polarization response, our metasurfaces feature large throughput efficiencies, approaching the theoretical maximum for a given NA. Our results pave the way for low-loss, high-efficiency and broadband optical computing and image processing with free-space metasurfaces.*

## Introduction

Image processing plays a key role in rapidly advancing technologies such as augmented reality, advanced driver assistance systems and biomedical imaging, and it is typically performed digitally. Despite their versatility, digital approaches are affected by several drawbacks, such as low operational speed and energy consumption, which are critical factors in several applications. These limitations can be overcome by developing analog components, optimized to perform specific calculations, which run in parallel to digital processors in order to enhance the overall speed and efficiency. Optics is the foremost approach for implementing such analog computations, due to the unmatched speed, low power consumption [1], [2], and ease of re-configurability into various network topologies [3]–[7]. For example, analog image processing is commonly performed via Fourier filtering techniques [8]–[10], whereby a 4F lens system is used to physically access the Fourier transform of an input image and filter different frequency components via spatially selective masks. Despite its simplicity, the required 4F configuration makes this approach inherently bulky and prone to alignment issues, and thus not well-suited for integrated devices. In recent years, optical computing has garnered renewed interest because the current nanotechnology tools have enabled the development of artificially engineered materials that allow to increase complexity [11] while maintaining compact footprints. In this context, metasurfaces – planarized ultra-thin artificial devices - have already been successfully employed for several tasks such as light wavefront shaping [12], thin polarization optics [13] and compact spectrometers [14].

Recently, it has been proposed [15], [16] that metasurfaces can be also used as compact devices to implement analog image processing without requiring to physically access the Fourier space, as schematized in Fig. 1a. Assume that an optical image is defined in the plane z = 0 by an intensity profile $I_{in}(x,y) = |\mathbf{E}_{in}(x,y)|^2$, where $\mathbf{E}_{in}(x,y) = E_{in}(x,y)\mathbf{e}$ is an electric field with polarization direction $\mathbf{e}$ and angular frequency $\omega = 2\pi c/\lambda = k_0 c$. Following standard Fourier optics [17], the image can be decomposed into a bundle of plane waves, each propagating along a direction identified by the polar and azimuthal angles $\theta$ and $\phi$ and with amplitude proportional to the Fourier transform $f_{in}(k_x, k_y) = \int dx dy \, e^{-i(k_x x + k_y y)} E_{in}(x,y)$, where $[k_x, k_y] = k_0 \sin\theta[\cos\phi, \sin\phi]$. In a standard imaging setup, these plane waves are collected and re-focused by a pair of lenses/objectives (not shown in Fig. 1a), effectively performing an inverse Fourier transform and re-creating the image at a different plane. Thus, any mathematical operation defined in the Fourier space can be performed by selectively filtering the bundle of plane waves originating from the image with respect to their propagation direction. This angle-selective filtering – typically hard to control with naturally available materials - can be implemented with suitably designed metasurfaces. By tailoring the scattering properties of the metasurface, the desired mathematical operation can be encoded into the metasurface transfer function.

An analog operation that has received much attention in the past years is the edge detection [18]–[29], whereby the edges of an input image $E_{in}(x,y)$ are enhanced with respect to homogeneous regions (Fig. 1a). This can be obtained by applying a spatial differential operator, such as the Laplacian operator $E_{out}(x,y) = (\partial_x^2 + \partial_y^2)E_{in}(x,y)$, which translates, in Fourier space, into a high-pass filter described by $f_{out}(k_x, k_y) = -(k_x^2 + k_y^2)f_{in}(k_x, k_y)$. Qualitatively, this operation can be obtained by a metasurface that suppresses plane waves propagating at small angles ($\theta \approx 0$), while progressively transmitting waves propagating at larger angles (Fig. 1a). In order to perform this operation on an arbitrarily polarized input 2D image, a metasurface must feature, for any azimuthal angle and for any polar angle below a certain maximum value $\theta \leq \theta_{max}$, a polarization-independent and isotropic transfer function of the form [19]

$$t(\theta, \phi) = \begin{pmatrix} t_{ss}(\theta, \phi) & t_{sp}(\theta, \phi) \\ t_{ps}(\theta, \phi) & t_{pp}(\theta, \phi) \end{pmatrix} = \begin{pmatrix} C \sin^2 \theta & 0 \\ 0 & C \sin^2 \theta \end{pmatrix} \quad (1)$$

where $C$ is an overall pre-factor and the subscripts denote s and p polarization (see also mathematical derivation in Supplementary Information, Supplementary Section S3). While in numerical experiments any transfer function described by Eq. 1 leads to second-order differentiation of a certain class of input images, in practical applications several figures of merit and key metrics need to be optimized to guarantee acceptable performance. To be able to process high resolution images without image distortion, the metasurface numerical aperture $\text{NA} \equiv \sin(\theta_{max})$ must satisfy $\text{NA} \geq k_{in,max}/k_0$, where $k_{in,max}$ is the largest wave vector in the Fourier decomposition of the input image. Thus, the ideal behavior described by Eq. 1 must remain valid up to very large polar angles. Moreover, low values of $|C| \ll 1$ are detrimental, since they strongly reduce the intensity of the processed image, resulting in large inefficiencies. Finally, while responses similar to Eq. 1 can be readily engineered at a given frequency, in many practical applications the input light might either have an unknown frequency or be polychromatic, and with a broad spectrum. Thus, it is highly desirable to use a metasurface for which the transfer function in Eq. 1 is maintained over a broad range of frequencies.

Several theoretical [18]–[22] and experimental [23]–[29] works have discussed different approaches to perform edge detection with metasurfaces under certain excitation/collection conditions. A common approach [19]–[21], [24], [28] relies on inducing a single optical mode in a nonlocal metasurface, leading to a Fano lineshape in the normal-incidence transmission spectrum, as sketched in Fig 1b. As a result, the normal-incidence transmission is zero at a single operational frequency ω (red dashed line in Fig. 1b). As the angle $\theta$ increases, the Fano lineshape spectrally shifts due to the metasurface angular dispersion, leading to the desired increase of the transmission at the fixed frequency ω, as sketched in Fig. 1c. While this approach has proven successful in realizing analog edge detection, its operational bandwidth is inherently limited. Indeed, the almost-zero transmission at normal incidence, necessary to suppress the low-spatial-frequency components, can be obtained only in a range of frequencies much smaller than the linewidth of the Fano resonance. The operational bandwidth can be increased by employing an optical mode with lower Q factor, in order to obtain a wider band of almost-zero transmission at normal incidence [28]. However, modes with larger linewidths also lead to a slower increase of the transmission as the angle θ increases. This reduces the transmission of the high-spatial-frequency components, effectively establishing a trade-off, common to most of the current literature, between spectral bandwidth and intensity throughput. Moreover, the nonlocal metasurfaces demonstrated experimentally so far feature either a $C_2$ [24] or $C_4$ [28] rotational symmetry, which introduce large azimuthal anisotropies and/or strong polarization asymmetries.

Approaches based on topological photonics [27], [29] can provide broader spectral bandwidths and better isotropy, but they typically require working in a cross-polarized reflection modality and at well-defined and large angles of incidence, which sets hard lower bounds on the footprint of the overall device and strongly limits the numerical aperture and practical operation. A largely different class of edge-detection devices has been recently proposed by placing Pancharatnam-Berry phase metasurfaces in the Fourier plane of a 4F system combined with two crossed polarizers [30]–[32]. This approach typically leads to large operational spectral bandwidths, since working in a 4F modality allows to effectively decouple the spectral and angular response of the metasurface. However, the 4F arrangement and the required polarizers strongly constrain the minimum footprint of this approach, and thus limit the possibility of miniaturization. We note that, as for any analog optical filtering technique that relies on coherent superposition of different plane waves, this approach to edge detection requires some degree of spatial coherence in the illumination. Recent works have proposed a way to overcome this limitation [33], [34], albeit at the expense of increased footprint and the need of digital postprocessing.

Overall, despite a few implementations discussed in the literature, a metasurface that can perform isotropic, broadband, polarization-independent, high-NA and high-efficiency edge detection without using a 4F arrangement, while keeping the footprint compact and the design feasible for fabrication, has not been demonstrated yet. In this context, it is worth emphasizing that image-processing approaches based on using a spatially-varying metasurface as a mask in a 4F system [26], [30]–[32] offer very limited possibilities in terms of miniaturization, since the footprint of the overall setup is largely dominated by the lens thicknesses and focal lengths. Moreover, it is important to stress that the efficiency of the edge-detection process, i.e., how the intensity of the output image compares to the intensity of the input image, has been overlooked in recent works, yet it is of paramount importance for practical applications.

In this work we propose and experimentally demonstrate a route based on dispersion engineering to realize edge-detection metasurfaces with optimal figures of merit discussed above. We showcase the potential of our design principle in a silicon-on-glass platform, demonstrating designs with isotropic responses, NAs larger than 0.35, operational bandwidths of 35 nm around a central wavelength of about 1500 nm, and operation for any input polarization – important requirements to facilitate the adoption of these devices in real-world application. In order to quantify the edge-detection performance, we introduce two metrics to assess the efficiency of our devices, quantifying in a more rigorous way the throughput and insertion loss of the metasurface –important benchmarks when comparing the performance of different designs. Importantly, we demonstrate that the efficiency experimentally achieved in our work is close to the efficiency of an ideal k-space filter with the same NA. While in the experiments shown here we focus on a design with a bandwidth of 35 nm (5 THz), the bandwidth can be increased by further engineering the metasurface dispersion. In the Supplementary Information (Supplementary Section S6), we numerically demonstrate a design where the bandwidth is increased two-fold, albeit at the expense of a slightly increased background.

## Results

**General Principle and Metasurface Design**

We first introduce the general concept behind our approach, and later discuss its implementation in a realistic design. Our recipe to obtain broadband, high-efficiency and high-NA edge-detection relies on engineering the band structure dispersion of a periodic nonlocal metasurface. Specifically, we engineer two different dispersive modes, denoted $\omega_+(\mathbf{k}_\parallel)$ and $\omega_-(\mathbf{k}_\parallel)$ (Fig. 1c), such that: (I) the frequency of the two modes are different but close at the $\Gamma$ point ($\mathbf{k}_\parallel = 0$), with $\Delta \equiv \omega_+(0) - \omega_-(0) > 0$, (II) the radiative linewidths of the modes $\omega_+(0)$ and $\omega_-(0)$ are comparable to or slightly larger than their detuning $\Delta$, and (III) the two modes shift spectrally towards different directions as $|\mathbf{k}_\parallel|$ increases, with the frequency $\omega_+(\mathbf{k}_\parallel)$ increasing and the frequency $\omega_-(\mathbf{k}_\parallel)$ decreasing. This behavior, schematically depicted in Fig. 1c, can be obtained for example at photonic crystal bandgaps. As shown schematically in Fig. 1d, conditions (I) and (II) result into a large band $B > \Delta$ of almost-zero transmission at normal incidence (solid blue line and orange-shaded area in Fig. 1d). The spectral width of this band is approximately double the linewidth of the two modes. Furthermore, due to condition (III), as the angle $\theta$ increases the transmission increases at all frequencies within the band B. Moreover, by ensuring that the dispersion of the two modes is strong enough that their overall spectral shifts are larger than their linewidths, the transmission within the band $B$ will rise to almost-unitary values for large angles $\theta$. This property ensures high transmission of large spatial-frequency Fourier components, resulting in higher intensities in the output images. Finally, while the previous guidelines can be applied to any periodic metasurfaces with arbitrary lattice symmetry, working with lattices with $C_6$ rotational symmetry leads to a polarization-independent response at normal-incidence, and it guarantees the largest possible degree of isotropy for tilted angles. We emphasize that a $C_6$ rotational symmetry does not automatically guarantee polarization-independent response at off-normal angles, which is instead crucial to achieve uniform edge detection independently of the polarization of the input image. Achieving a full-angle polarization-independent response requires to further engineer the dispersion of the metasurface to ensure that the relevant optical modes have a similar coupling to s- and p-polarized waves.

Remarkably, we show that all these seemingly stringent requirements can be achieved within a relatively simple metasurface platform composed of a photonic crystal slab over a transparent substrate (Fig. 2). In this work we consider silicon metasurfaces operating in the near-infrared (NIR), but the concept can be readily generalized to any spectral region. Figure 2a shows the unit cell of the proposed design, which consists of a triangular lattice of air holes etched into a silicon slab (relative permittivity $\varepsilon_r$=11.90) and placed on glass ($\varepsilon_r$=2.31) for mechanical support. The device design is fully characterized by three parameters: the lattice constant $a$, the thickness $H$ and the radius of the holes $R$. We fix the lattice constant to $a$ = 924 nm in order to achieve operational wavelengths close to 1500 nm, while we vary the slab thickness $H$ and the hole radius $R$ to optimize the band structure and meet the three conditions described above. All simulations were performed numerically with a commercially available software (Ansys HFSS). Figure 2b shows the normal-incidence transmission amplitude, for fixed $R$ = 265 nm, versus impinging wavelength and for various thicknesses $H$. The transmission spectra are dominated by two modes, whose distance can be readily controlled by $H$. Following conditions I and II described above, we select $H$ = 270 nm (dashed-dotted line in Fig. 2b), where the spectral detuning of the two modes is comparable to their linewidths. Figures 2c shows, for this optimized device, the simulated p-polarized transmission amplitude $|t_{pp}|$ versus wavelength and for increasing values of the polar angle $\theta$. As clear from the color plot, the two modes follow the targeted behavior (see also Figs. 1c-d) diverging from each other up to large angles $\theta_{max} \approx 26°$. This feature results in a bandwidth of about 35 nm (5 THz), centered approximately around 1500 nm, over which the desired Laplacian-like transfer function is obtained. We note that the spectral bandwidth could be increased by working with optical modes that are further detuned at normal incidence, albeit at the expense of a larger background at the central frequency. As an example, in the Supplementary Information (Supplementary Section S6) we numerically demonstrate a design with slightly different values of $R$ and $H$, which features a bandwidth of 10 THz.

The quality of the Laplacian-like response is further highlighted in Fig. 2d, where we show vertical cut lines at selected wavelengths corresponding to the color-coded vertical lines in Fig. 2c. For all wavelengths, $|t_{pp}(\theta)|$ increases almost monotonically with $\theta$, reaching values above 0.9 for $\theta \approx 30°$. The s-polarized transmission amplitude, shown in Figs. 2(e-f), displays a similar behavior although with some nonidealities. At $\theta = 0°$ the two transmission zeros occur at the same wavelengths as for p-polarization, due to the triangular lattice symmetry. For $\theta > 0°$, the lower-wavelength mode shifts towards shorter wavelengths, as required. The higher-wavelength mode, instead, is almost dispersionless for small values of $\theta$ and it shifts towards shorter wavelengths as $\theta$ increases. Therefore, the associated transmission zero crosses through the operational band. As a result, for certain wavelengths the transmission amplitude does not increase monotonously with $\theta$, but it instead features a zero for intermediate values of $10° < \theta < 20°$ (Fig. 2f). Despite this potential issue, we note that all transfer functions in Fig. 2f display the desired high-pass-filter behavior, suppressing low spatial frequencies while promoting high frequencies. As we demonstrate experimentally in the following, this sub-ideal behavior for s-polarization does not introduce any practical detrimental effect in the edge-detection functionality. Moreover, the transmission amplitudes reach values of almost 1 for $\theta \approx 20°$ (Fig. 2f), and they remain almost flat for larger angles, further highlighting the large efficiency and NA of this device.

In order to verify the edge-detection performance of this metasurface, and in particular its isotropy, we calculated the angle-dependent transfer functions $t_{pp}(k_x, k_y)$ and $t_{ss}(k_x, k_y)$ for each of the five wavelengths considered in Figs. 2(d,f). The amplitudes of the transfer functions [Figs. 3(a-e) for p-polarization and Figs. 3(f-j) for s-polarization] show that, up to $\theta \approx 21^o$ (NA = 0.35, dashed circles in the transfer functions in Fig. 3), the device response is almost fully isotropic with respect to the azimuthal angle $\phi$. The phases of the transmission amplitudes, shown in the Supplementary Information (Supplementary Section S2.1), are almost independent of $\theta$ and $\phi$, as required for a proper implementation of the Laplacian operation. The cross-polarized transfer functions, $t_{sp}(k_x, k_y)$ and $t_{ps}(k_x, k_y)$, shown in the Supplementary Information (Supplementary Section S2.1), have vanishing amplitudes up to NA = 0.35. In order to verify the edge detection functionality, we numerically calculated the output images created by this metasurface for different input wavelengths, using the CUNY logo shown in Fig. 1a as the input image. The pixel size of the input image is rescaled such that the lateral extent of the logo is equal to about $80\lambda$ (~120 $\mu m$). In order to readily estimate the edge detection efficiency (see discussion below), we set the peak intensity of the input image equal to 1. To obtain the output image (see Supplementary Information, Supplementary Section S3 for the detailed derivation), we calculate the full polarization-dependent plane-wave expansion of the input image, and then calculate the filtering effect of the metasurface by including both co-polarized ($t_{ss}$ and $t_{pp}$) and cross-polarized ($t_{sp}$ and $t_{ps}$) complex transfer functions. Moreover, we assume that the input image is created by an unpolarized electric field. The output images in Figs. 3(k-o) confirm that the metasurface filtering results into sharp edges and a suppression of the homogeneous background at multiple wavelengths (each plot is calculated for the wavelength reported on top the corresponding column). Thanks to the metasurface isotropy, the intensity and appearance of the edges are quite uniform and independent of the edge direction. In order to quantitatively assess the efficiency of our edge-detection metasurface, it is necessary to quantify how the intensity of the filtered image compares with the intensity of the input image. This metric is of crucial importance for the implementation of these devices in real-world applications because it dictates, for example, how much the integration time of a camera needs to be increased in order to have the same signal level in the output and input image. To this aim, we introduce two different metrics. First, we consider the *peak efficiency*, defined as the ratio $\eta_{\text{peak}} \equiv \max(I_{\text{out}})/\max(I_{\text{in}})$ between the peak intensities in the output and input images. The advantage of this metric is that it can be easily estimated. Since we have set $\max(I_{\text{in}}) = 1$, the efficiency $\eta_{\text{peak}}$ is equal to the upper limit of each colorbar in Figs. 3(k-o), and it ranges between 7% and 9%. However, due to its definition, the value of $\eta_{peak}$ is typically determined by the highest-intensity pixels in the images (corresponding to small regions with very large derivatives), and it cannot correctly quantify the global efficiency. We therefore introduce also the *average efficiency* $\eta_{\text{avg}} \equiv \text{avg}\left(I_{\text{out}}^{\text{edge}}\right)/\max(I_{\text{in}})$, where $\text{avg}\left(I_{\text{out}}^{\text{edge}}\right)$ is the average intensity of the output image calculated only in narrow regions surrounding the expected positions of all edges. For the images in Figs. 3(k-o), $\eta_{avg}$ varies between 1.5% and 3% depending on the wavelength. As we demonstrate in the Supplementary Information, these efficiencies are fairly close to the maximum theoretical value obtainable with an ideal edge-detecting device with the same NA. We emphasize that such high efficiencies are achieved due to the fact that our metasurface provides a Laplacian-like response for both polarizations, and with large values of transmission amplitudes at large angles.

**Experimental Verification**

After having numerically verified the performance of the proposed metasurface design, we now experimentally demonstrate its operation. We fabricated the metasurface in Figs. 2-3 by using amorphous silicon deposited on a glass substrate (see Methods for details on fabrication), and its transfer functions were measured with a custom-built setup (see Methods for details on the setup). The experimental measurements are shown in Figs. 2(g-j) side-by-side to the corresponding simulated results, and they show excellent agreement with the calculations. Specifically, for p-polarized excitation the transmission amplitude versus wavelength and $\theta$ (Fig. 2g) clearly shows the occurrence of the two transmission zeros at normal incidence, and their diverging spectral shift for increasing values of $\theta$. Similarly, the measured s-polarized transmission amplitude (Figs. 2i-j) agrees well with the simulated plots (Figs. 2e-f), and it displays the expected global increase of the transmission amplitude versus $\theta$ (Fig. 2j). All measurements and simulations in Fig. 2 are performed for a fixed value of the azimuthal angle $\phi = 0$. The experimental data for $\phi = 30^o$ are available in the Supplementary Information (Supplementary Section S2.2).

Next, we demonstrate experimentally the edge-detection functionality of our device. To create the input image, we used a target with the shape of our institution logo, obtained by depositing a 200-nm-thick layer of chromium on a glass substrate and then etching the desired shape after an electron beam lithography process. In our setup (see Fig. 4a and additional details in Supplementary Information, Supplementary Section S1), the target is illuminated by a collimated beam, which acts as a wide-field illumination, and the image scattered by the target is collected by a NIR objective (Mitutoyo, 50X, NA = 0.42) and relayed on a near-infrared camera with a tube lens with a $f$=15cm focal length. To perform edge detection, the metasurface is placed between the objective and the target, at a distance of few hundreds microns from the target. In order to correctly quantify the efficiencies $\eta_{\text{peak}}$ and $\eta_{\text{avg}}$, in all measurements (with and without the metasurface) we normalize the counts read by the camera by the camera integration time and the power impinging on the target.

We performed a first experiment with almost-unpolarized (degree of polarization = 10%) and narrowband (FWHM ≈ 5 nm) illumination, which is obtained by filtering the output of a supercontinuum laser with a commercial tunable filter. Figure 4b shows the output image obtained without the metasurface (top-left plot), together with several output images obtained with the metasurface in front of the target and at different input wavelengths, swept across the operational bandwidth determined by Figs. 2(g-j). The illumination wavelength (in nanometers) is reported in the bottom-right corner of each color plot of Fig. 4b. As clear from these measurements, high-quality edge detection can be obtained for any wavelength between 1452 nm and 1485 nm. The extent of this bandwidth matches very well the simulations in Fig. 3. Moreover, for all wavelengths between 1452 nm and 1485 nm the efficiencies $\eta_{peak}$ and $\eta_{avg}$ reaches values larger than 5% and larger than 1%, respectively, experimentally confirming that our metasurface can perform broad-bandwidth edge detection while maintaining a very large throughput. As mentioned in the previous section, values of $\eta_{peak} \geq 5\%$ are very close to the peak efficiency obtainable with an ideal k-space filter performing the Laplacian operator with a fixed NA=0.35 ($\eta_{peak}^{(ideal)} = 8\% - 9\%$) (Supplementary Information, Supplementary Section S4). Thus, the experimental efficiency of our metasurface is very close to the upper theoretical bound. The edge-detection performance is further confirmed by the horizontal-cut intensity distributions shown in Fig. 4c (corresponding to the white dashed line in top-left plot of Fig. 4b). High-intensity peaks at the expected edge positions are obtained, surrounded by an almost-zero and uniform background. As expected from the second-order differentiation of a step-like function, each edge results in a pair of high-intensity peaks in the output images. From a practical point of view, the presence of these two peaks (as opposed to a single peak) can be useful to better identify the spatial position of the detected edge. Besides the two high-intensity peaks, some additional weaker peaks and noise are visible in the output images (see also Fig. 4c). As explained in more detail in the Supplementary Information (Supplementary Section S5), the additional weaker peaks are due to the fact that, in the input image, the intensity is not exactly constant inside (or outside) the CUNY logo. Instead, weak intensity fluctuations are present due to diffraction at the metallic aperture used to create the input image (see also Fig. S8 in Supplementary Information). Since our metasurface performs high-efficiency differentiation on the whole input image, these spatial intensity fluctuations in the input image will be differentiated as well, resulting in additional sets of weaker peaks in the output image. From a practical point of view, the presence of these additional weaker peaks does not limit the capability of detecting the position of the main edges of the image, because the intensity of each peak in the output image remains proportional to the square of the spatial derivative (at the same position) of the input image. In our experiment, the peaks in the output images corresponding to the main edges are about 4x more intense than the peaks corresponding to the weak intensity fluctuations.

Finally, we experimentally verify the quality of the edge detection under a broad-band input. A broad excitation spectrum, extending from 1450 nm to 1485 nm (Fig. 5a), was obtained by filtering the output of a supercontinuum laser with a custom-built pulse shaper. Fig. 5b shows the unfiltered image (i.e., without the metasurface), while Figs. 5(c-h) show the filtered image for different polarizations of the input illumination (as described in the figure caption). Our experimental results confirm that the output images feature very sharp and high-contrast edges also for very broad excitations. Remarkably, these measurements confirm that the quality and uniformity of the edges is essentially independent of the input polarization. This confirms that, despite the sub-ideal features for *s*-polarized excitation (Figs. 2e and 2i), our metasurface is still capable of performing high-quality edge detection for any input polarization. Moreover, even for broadband and arbitrarily polarized input, the efficiency remains quite high, with $\eta_{\text{peak}} \geq 3.5\%$ and $\eta_{\text{avg}} \approx 1\ \%$ for all images in Fig. 5(c-h). The quality of the edges and the clear background suppression can also be appreciated by the horizontal cuts in Fig. 5i, which show the intensities of the filtered images (at the location marked by the horizontal white line in Fig. 5b) for the six different polarizations. In the Supplementary Information (Supplementary Section 2.3) we provide similar measurements for other target shapes.

## Discussion

In this work, we have theoretically proposed and experimentally demonstrated an approach to design metasurfaces performing analog edge detection over a broad bandwidth of input frequencies and for arbitrary polarization inputs, while simultaneously maintaining high efficiency, high NA and excellent isotropy. Differently from conventional methods that rely on the angle-dependent spectral shift of a single optical mode, our approach relies on accurately engineering the angular dispersion of two different modes in a transversely invariant nonlocal metasurface. We have shown that this method provides enough flexibility and degrees of freedom to simultaneously optimize all the relevant figures of merit. Despite its apparent complexity, this approach can be readily implemented in different platforms and materials. In particular, we have experimentally implemented our approach in a silicon-on-glass platform, demonstrating a simple single-layer metasurface with isotropic responses up to large numerical apertures (NA > 0.35), and with an operational bandwidth of 35 nm (5 THz) around a central wavelength of about 1500 nm. We experimentally verified that this device performs high-quality and high-efficiency second-order differentiation for any input polarization state and for any wavelength within the bandwidth. Moreover, in order to fully and quantitively assess the performance of the device, we introduced two different efficiency metrics, to quantify how the peak and average intensity of the edges compares to the intensity of the input image. Thanks to its dual-polarization response and to the large transmission values achieved at large angles, our device can achieve very large efficiencies for both monochromatic and broadband inputs, and for any input polarization. In particular, the experimentally measured efficiencies demonstrated in this work are very close to the maximum theoretical efficiency achievable with a passive k-space filter with a given

NA, and they are only a factor of ~2x smaller than the efficiency that would be obtained by applying the exact Laplacian operation to the input images.

We emphasize that the spectral bandwidth demonstrated here (35 nm around 1470 nm, corresponding to 5 THz) can be increased by further engineering the dispersion of the metasurface, i.e., by further optimizing the position, linewidth, and dispersion of the two optical modes. In particular, larger operational bandwidths can be obtained if the spectral detuning between the two optical modes considered in Fig. 2 (see also the range indicated by $\Delta$ in Fig. 1c) is increased, provided that the quality factor of each mode is simultaneously decreased in order to keep the normal-incidence transmission low. As an example, in the Supplementary Information (Supplementary Section S6) we show numerical calculations of an additional optimized design, obtained by slightly tweaking the geometry in this work, which features a bandwidth of 10 THz, i.e., twice the one of the device considered in Figs. 2-5. Moreover, this dispersion engineering approach demonstrated here can be expanded to scenarios involving more than two optical modes, which allow a larger operational bandwidth and additional flexibility in engineering the angle-dependent response, albeit at the expense of increased design complexity.

Our results demonstrate that it is possible to design and realize simple single-layer metasurfaces to perform analog image computation in realistic scenarios where, for example, the input polarization or frequencies are unknown, and when large insertion loss cannot be tolerated. Moreover, the simplicity of the proposed design makes it amenable to mass manufacturing, an important requirement for future commercialization of these devices. We anticipate that further improvement of the proposed design can lead to even larger operational bandwidths and efficiencies. Moreover, we expect that more complex dispersion engineering, such as controlling the position and dispersion of additional zeros and poles, may lead to more sophisticated transfer functions and advanced functionalities.

## Methods

**Sample Fabrication**

The samples were fabricated with a standard top-down lithographic process. Glass coverslips (25 x 75 x 1 mm, Fisher Scientific) were used as transparent substrates. The substrates were cleaned by placing them in an acetone bath inside an ultrasonic cleaner, and later in an oxygen-based cleaning plasma (PVA Tepla IoN 40). After cleaning, a layer of 270 nm of amorphous silicon (α-Si) was deposited via a plasma-enhanced chemical vapor deposition (PECVD) process. A layer of E-beam resist (ZEP 520-A) was then spin-coated on top of the samples, followed by a layer of an anti-charging polymer (DisCharge, DisChem). The desired photonic crystal pattern was then written with an electron beam tool (Elionix 50 keV). After ZEP development, the pattern was transferred to the underlying silicon layer via dry etching in an ICP machine (Oxford PlasmaPro System 100). The resist mask was finally removed with a solvent (Remover PG). SEM pictures of the final device are shown in the Supplementary Information, Fig. S1.

**Optical Characterization**

The transmission amplitudes shown in Figs. 2(g-j) were performed with a custom-built setup described in more detail in the Supplementary Information (Section S1). The sample was mounted on two different rotation stages to control the polar angle $\theta$ and the azimuthal angle $\phi$. A broadband supercontinuum laser (NKT, SuperK) was filtered via a commercial narrowband filter (Photon, LLTF Contrast) and then injected into the setup via a fiber. The laser was weakly focused on the metasurface via a lens with f = 20 cm focal length. The transmitted signal was collected and re-collimated on the other side of the sample by an identical lens. Two identical germanium powermeters (Thorlabs, S122C) were used to measure the transmission level through the metasurface. A linear polarizer placed before the beamsplitter was used to polarize the incoming beam along either x or y, which correspond, respectively, to p- and s-polarization for any value of $\theta$ and $\phi$. The transmission amplitudes shown in Figs. 2(g-j) were then obtained by sweeping the angle $\theta$ and the input wavelength and recording the powers measured by the powermeters.

The imaging experiments shown in Figs. 4 and 5 of the main paper were performed with the setup shown in Fig. 4a. The illumination was provided by the same supercontinuum source used in the setup described in the previous paragraph. For the measurements in Fig 4(b-c), the broadband source was filtered by the same narrow band filter used for the transmission measurements. For the measurements in Fig. 5, the output of the supercontinuum laser was filtered with a custom-built pulse shaper that allows to continuously tune the linewidth and central wavelength of the input spectrum.

# Data availability

All data that support the findings of this work are shown in the main text and Supplementary Information. The experimental data generated in this study and shown in Figs. 2, 4 and 5 have been deposited in the repository https://doi.org/10.5281/zenodo.10004018.

# Code availability

All electromagnetic simulations [Figs. 2(b-f) and Figs. 3(a-j)] were performed using a commercially available electromagnetic solver (Ansys HFSS), and they can be readily reproduced by using the geometrical parameters provided in the text. The simulated image processing results in Figs. 3(k-o) were obtained with a custom-made Matlab script that implements numerically the formulas shown in Supplementary Information (Supplementary Section S3).

## Acknowledgements

This work was supported by the Air Force Office of Scientific Research MURI program and the Simons Foundation. The authors would like to thank Dr. Dmitriy Korobkin for building the pulse shaper used for the experiment discussed in Fig. 5.

## Author Contributions Statement

All authors conceived the idea and the corresponding experiment. A.Ar. performed numerical simulations and optimization, and he conducted the numerical image analysis with assistance from M.C.. M.C. fabricated the devices and performed the experimental measurements together with S.S.. A.Al. supervised the project. All authors analyzed the data and contributed to writing the manuscript.

## Competing Interests Statement

The authors declare no conflicts of interest.

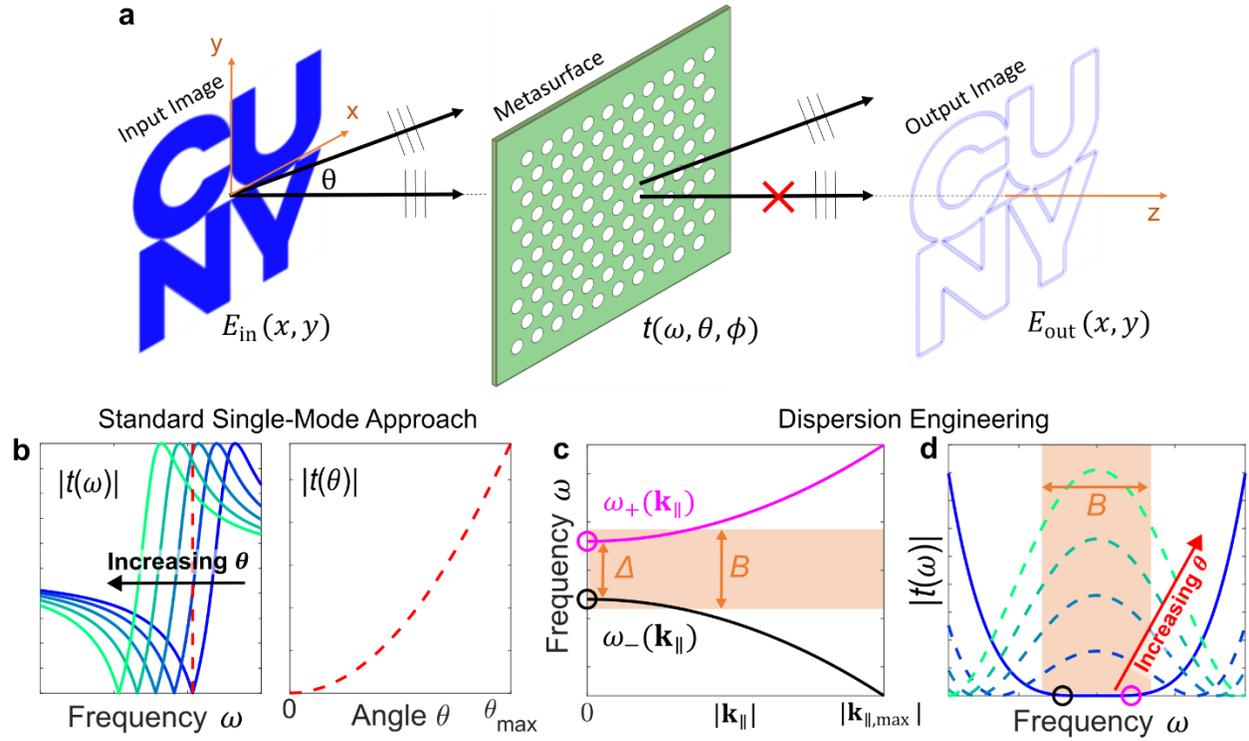

**Figure 1. Two-dimensional image differentiation using metasurfaces**. (a) Schematic of using a metasurface to implement the Laplacian operation on an input image. (b) Sketch of the conventional approach to achieve the Laplacian response at a single frequency. Left: the transmission amplitude $|t(\omega)|$ of a single-mode Fano lineshape spectrally shifts as a function of the incident angle. Right: at a single frequency (marked by the vertical dashed line in left part), the transmission amplitude versus angle ($|t(\theta)|$) displays the required Laplacian behavior. (c-d) Schematic of the concept demonstrated in this work. (c) Targeted dispersion relation, consisting of two modes $\omega_+(\mathbf{k}_\parallel)$ and $\omega_-(\mathbf{k}_\parallel)$ whose frequencies shift towards opposite directions as the in-plane wave-vector increases. (d) Resulting angle-dependent transmission amplitude, displaying a large bandwidth $B$ (shaded area) with almost-zero transmission at normal incidence.

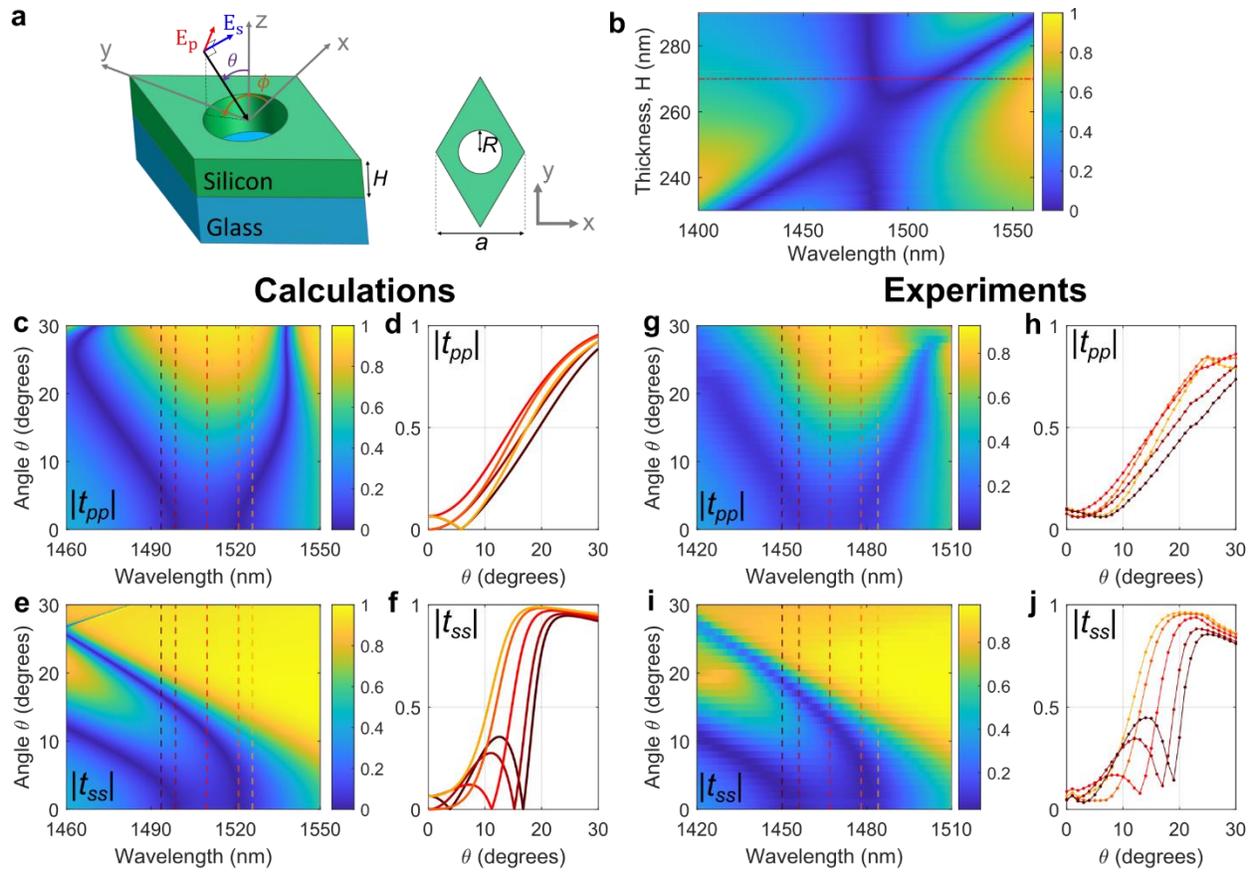

**Figure 2. Geometry, numerical simulations, and experimental measurements.** (a) Unit cell of the metasurface considered in this work. All devices considered in this work have lattice constant $a$ = 924 nm. (b) Numerically calculated normal-incidence transmission spectra for devices with fixed radius $R$ = 265 nm and varying thickness $H$. (c) Numerically calculated magnitude of the p-polarized transmission coefficients, $|t_{pp}|$, versus the incident angle $\theta$ and wavelength, for a device with $R$ = 265 nm and $H$ = 270 nm (red dashed-dotted line in panel b) and for $\phi = 0$. (d) Calculated $|t_{pp}|$ versus θ for five increasing wavelengths $\lambda_1$, $\lambda_2$, $\lambda_3$, $\lambda_4$, $\lambda_5$, denoted by color-coded vertical lines in panel c. (e-f) Same as panels c-d, but for the s-polarized transmission coefficient $|t_{ss}|$. (g-j) Same as panels c-f, but experimentally measured (see text and Supplementary Information for additional details).

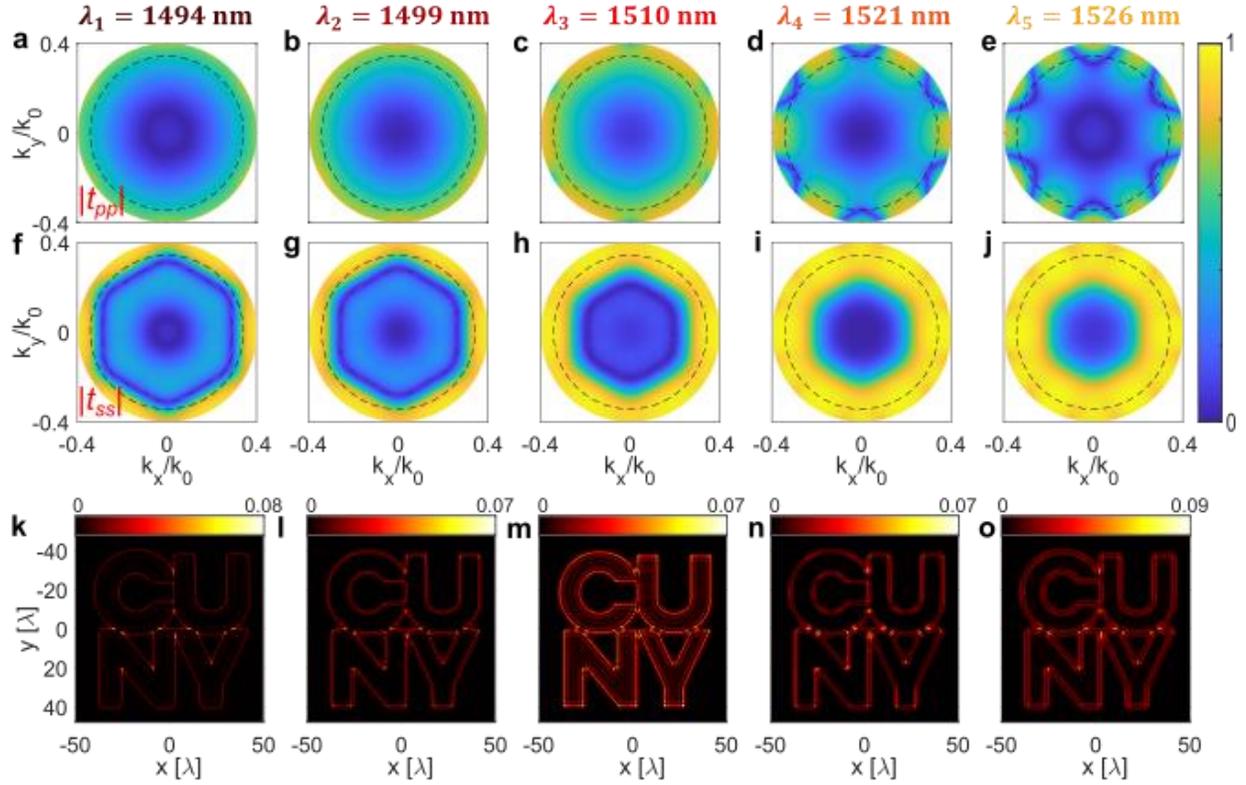

**Figure 3. Numerical calculations of transfer functions and edge detection** (a-e) Absolute value of the p-polarized transfer function of the device considered in Figs. 2(c-f), versus the in-plane wave-vector and for the five different wavelengths $\lambda_1 - \lambda_5$ considered in Figs. 2(c-f), as indicated above of each panel. The dashed circles denote NA = 0.35. (f-j). Same as for panels (a-e), but for the s-polarized transfer function. (k-o) Calculated output images for each wavelength and for unpolarized excitation. The input image is the CUNY logo shown in Fig. 1a, where the high-intensity areas (the inner part of the letters) have intensity equal to 1, while the lower-intensity areas have intensity equal to zero. For each of the panels (k-o), the efficiency $\eta_{peak}$ (defined in the text) can be obtained from the upper limit of the corresponding colorbar.

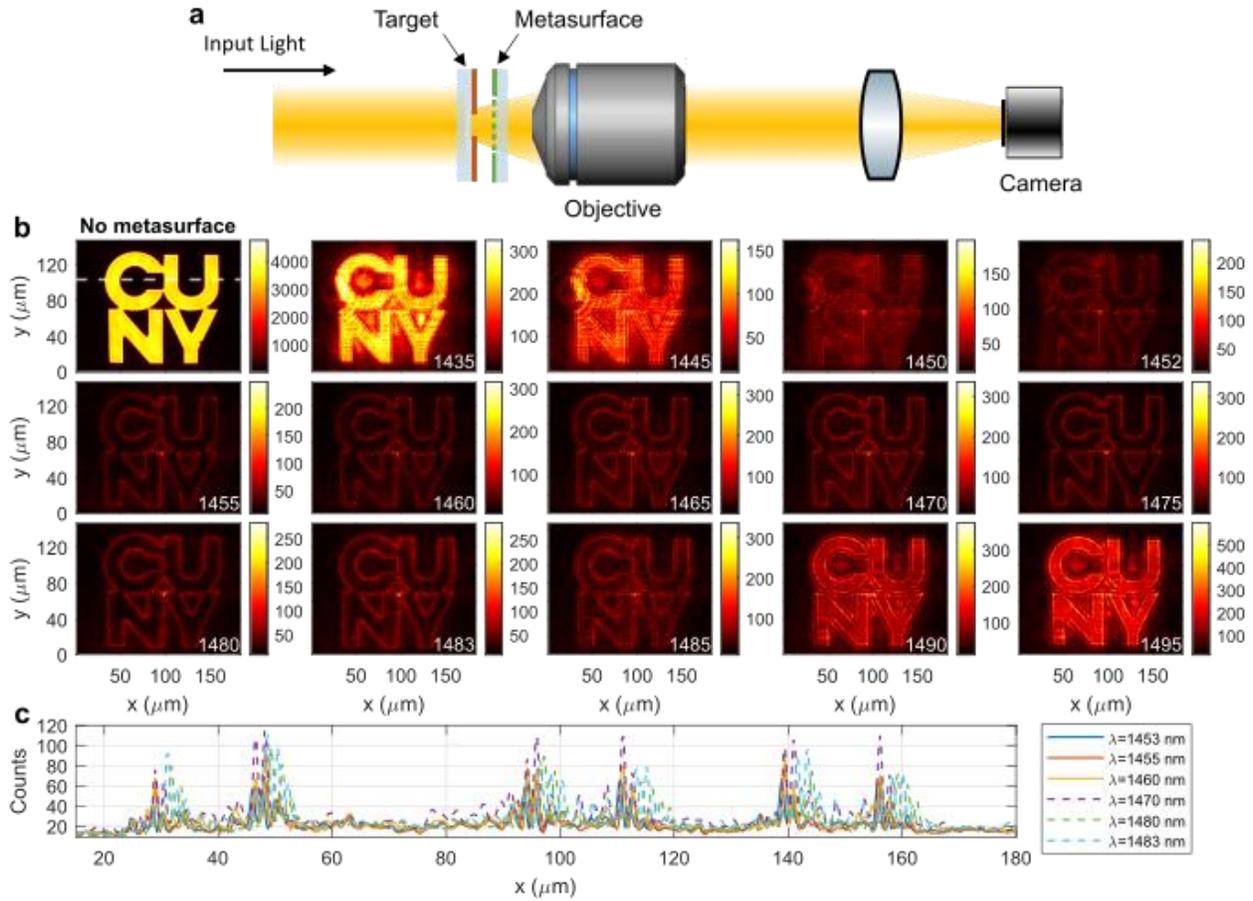

**Figure 4. Experimental edge detection with unpolarized narrow-band input.** (a) Schematic of the setup used for the imaging experiments. (b) Top-left plot: Unfiltered image, obtained by removing the metasurface from the setup in panel a. All other colorplots show the output images obtained when the metasurface is placed in front of the target, for different impinging wavelengths (reported at the bottom-right corner of each plot). (c) Horizontal cuts of selected wavelengths from panel b (as indicated in the legend), corresponding to the vertical position denoted by the white dashed line in the top-left plot of panel b. For all measurements (with and without metasurface), the counts recorded by the camera have been normalized by the camera integration time and the power impinging on the target. The illumination is almost unpolarized (measured degree of polarization ≈ 10 %).

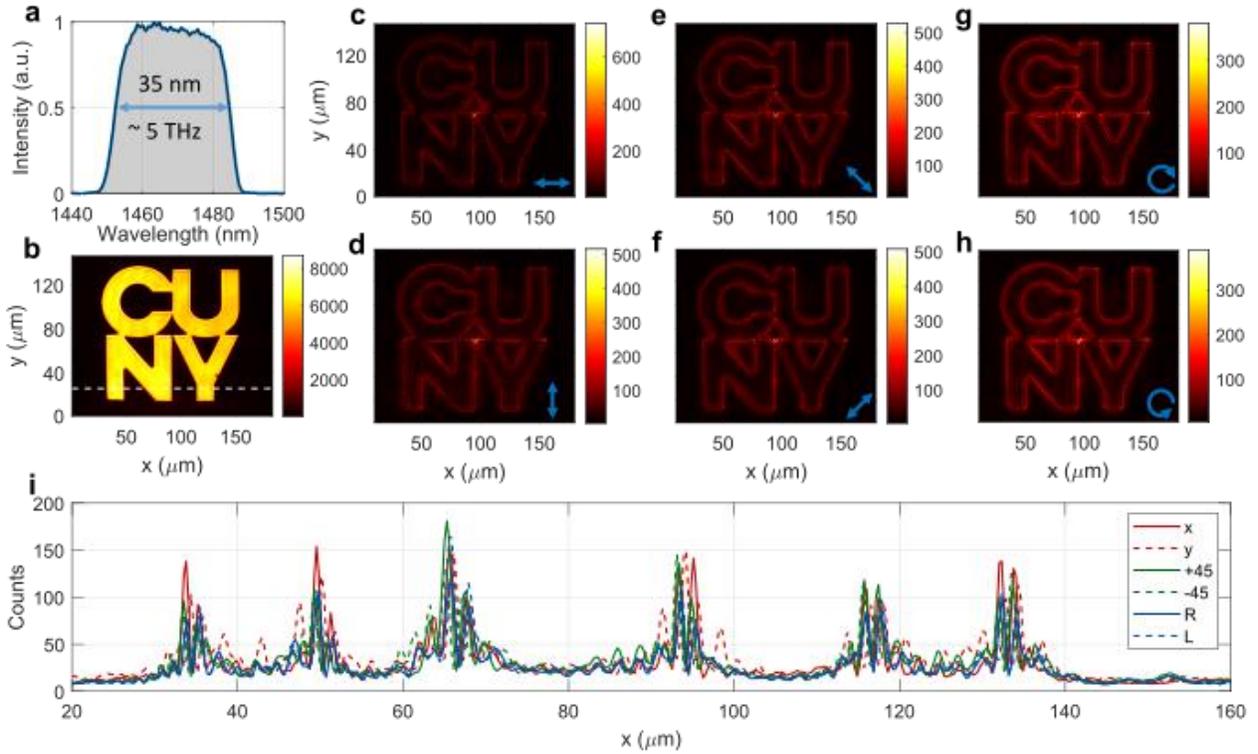

**Figure 5. Experimental edge detection with polarized broad-band input.** (a) Spectrum of the excitation. (b) Unfiltered image. (c-h) Output images when the metasurface is placed in front of the target and for six different polarizations of the input light: linearly polarized along (c) x, (d) y, (e) the x-y diagonal, (f) the x-y anti-diagonal, or (g) right and (h) left circularly polarized. (i) Horizontal cuts of the plots in panels (c-h) (as indicated in the legend), corresponding to the vertical position denoted by the white dashed line in panel (b). For all measurements (with and without metasurface), the counts recorded by the camera have been normalized by the camera integration time and the power impinging on the target.

# Table of Contents

# S1. Optical Characterization

The transmission amplitudes shown in Figs. 2(g-j) of the main text were performed with the custom-built setup shown in Fig. S1a. The sample was mounted on two different rotation stages, a motorized one (Thorlabs, HDR50) to control the polar angle $\theta$, and a manual one to control the azimuthal angle $\phi$. A broadband supercontinuum laser (NKT, SuperK) was filtered via a commercial narrowband filter (Photon, LLTF Contrast) and then injected into the setup via a fiber. The filtered laser has a linewidth of approximately 5 nm (see Fig. S1b). The laser was weakly focused on the metasurface via a lens (L1) with f = 20 cm focal length. The transmitted signal was collected and re-collimated on the other side of the sample by an identical lens (L2). Two identical germanium powermeters (Thorlabs, S122C), P1 and P2, were used to measure the transmission level through the metasurface. A beamsplitter (BS), placed before the excitation lens L1, was used to redirect approximately 50% of the laser power to the photodiode P2. A linear polarizer placed before the beamsplitter was used to polarize the incoming beam along either x or y, which correspond, respectively, to p- and s-polarization for any value of $\theta$ and $\phi$. A second polarizer after lens L2 was used to select the polarization of the collected beam. The transmission amplitudes shown in Figs. 2(g-j) were then obtained by sweeping the angle $\theta$ and the input wavelength and recording the powers measured by P1 and P2. An additional calibration run was taken without the metasurface, to account for the exact splitting ratio of the BS and for discrepancies between the two powermeters.

The imaging experiments shown in Figs. 4 and 5 of the main paper were performed with the setup shown in Fig. 4a. The illumination was provided by the same supercontinuum source used in the setup in Fig. S1a. For the measurements in Fig 4(b-c), the broadband source was filtered by the same narrow band filter used for the transmission measurements. For the measurements in Fig. 5, the output of the supercontinuum laser was filtered with a custom-built pulse shaper that allows to continuously tune the linewidth and central wavelength of the input spectrum.

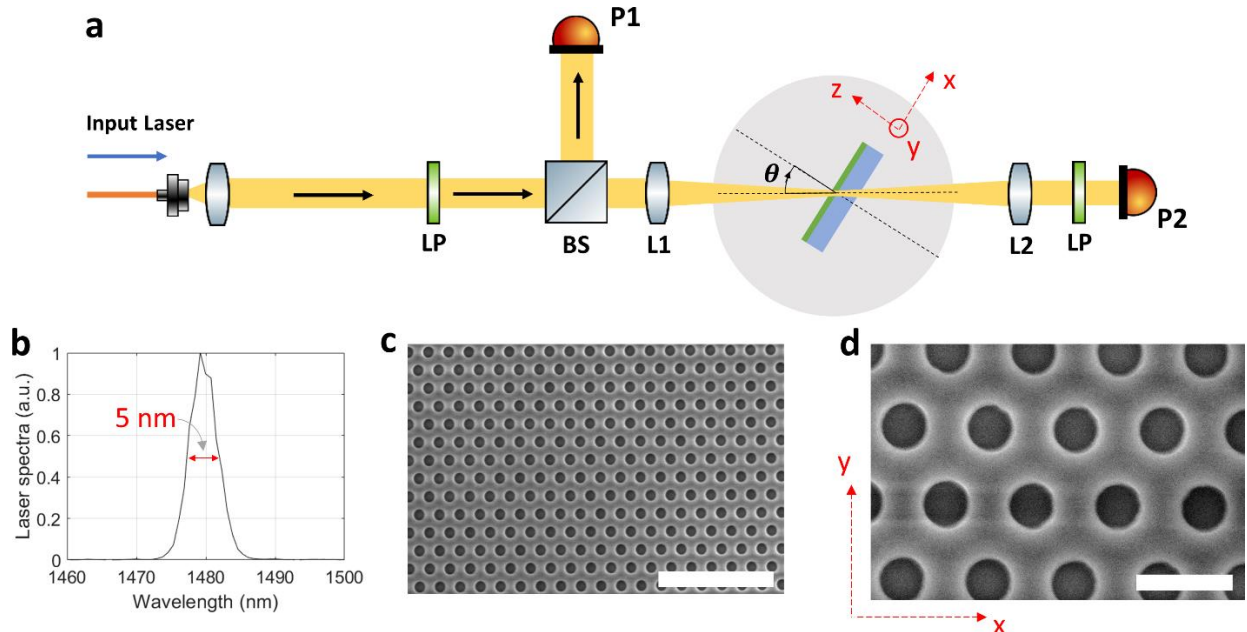

**Figure S1.** (a) Schematic of the setup used to perform the angle- and wavelength-dependent transmission measurements. LP = Linear polarizer, BS = beamsplitter, L1, L2 = lenses, P1, P2 = photodiodes. See text for additional details. (b) A representative spectrum of the laser used for the measurements shown in Figs. 2(g-j) and Fig. 4 of the main text, centered at 1480 nm. The laser has a linewidth of about 5 nm. (c-d) SEM pictures of the fabricated device, at different scales. Scale bars: 5 µm in panel c and 1 µm in panel d.

## S.2 Additional numerical and experimental data

### S2.1 Transfer functions

In Fig. 3 of the main text we showed the absolute values of the co-polarized transfer functions, $|t_{ss}|$ and $|t_{pp}|$, for five different wavelengths. The full co- and cross-polarized complex transfer functions, for the same set of wavelengths, are shown in Figs. S2 and S3. As mentioned in the main text, the magnitude of the cross-polarized transfer functions $|t_{ps}|$ and $|t_{sp}|$ (Figs. S2c and S2d, respectively) are very small within the numerical aperture NA = 0.35 (dashed circles n in each plot). The phases of $t_{pp}$ and $t_{ss}$ (Figs. S3a and S3b, respectively) are quite uniform upon variation of the azimuthal angle $\phi$, confirming the excellent isotropy of our devices. The phases are also fairly constant upon variations of the polar angle $\theta$, except for jumps that occur when the corresponding amplitude becomes zero (compares Figs. S3(a-b) with Fig. S2(a-b)). The phases of the cross-polarized transfer functions (Figs. S3(c-d)) show a more convoluted behavior. However, due to the very low values of the corresponding magnitudes, they play a negligible role in the overall imaging.

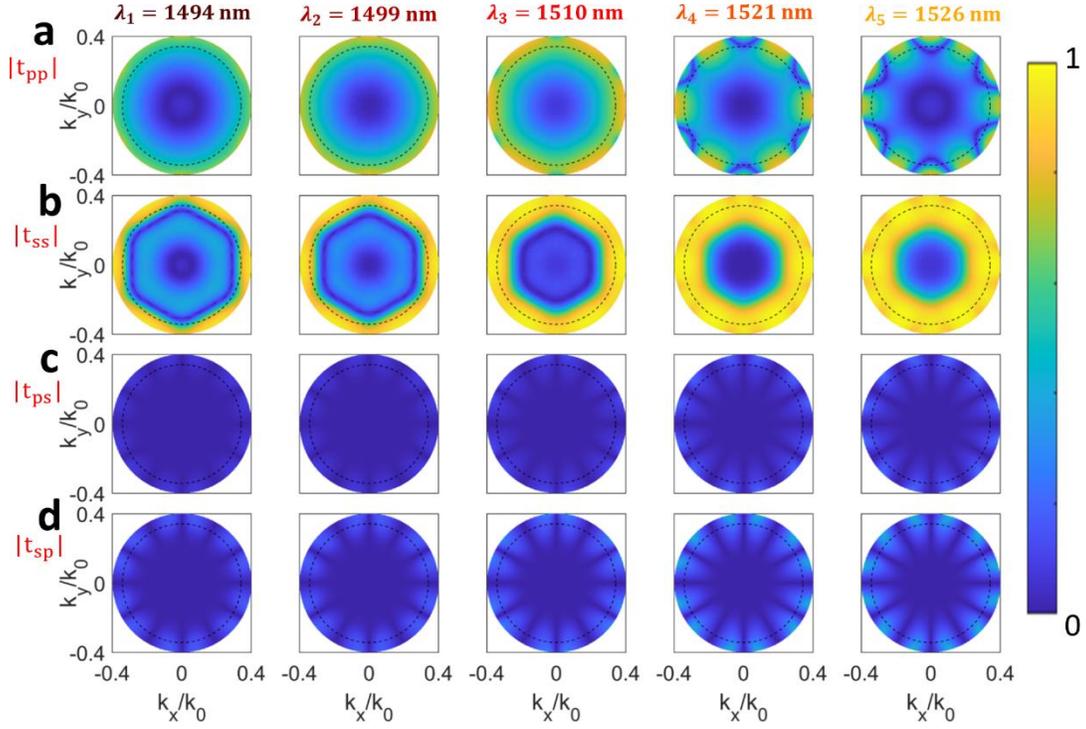

**Figure S2.** Calculated magnitude of the complex co- and cross-polarized transfer functions for the five wavelengths considered in Fig. 3 of the main text. Each column corresponds to a different wavelength, reported on top of the column. (a) Co-polarized transfer functions $|t_{pp}|$. (b) Co-polarized transfer functions $|t_{ss}|$. (c) Cross-polarized transfer functions $|t_{ps}|$. (d) Cross-polarized transfer functions $|t_{sp}|$.

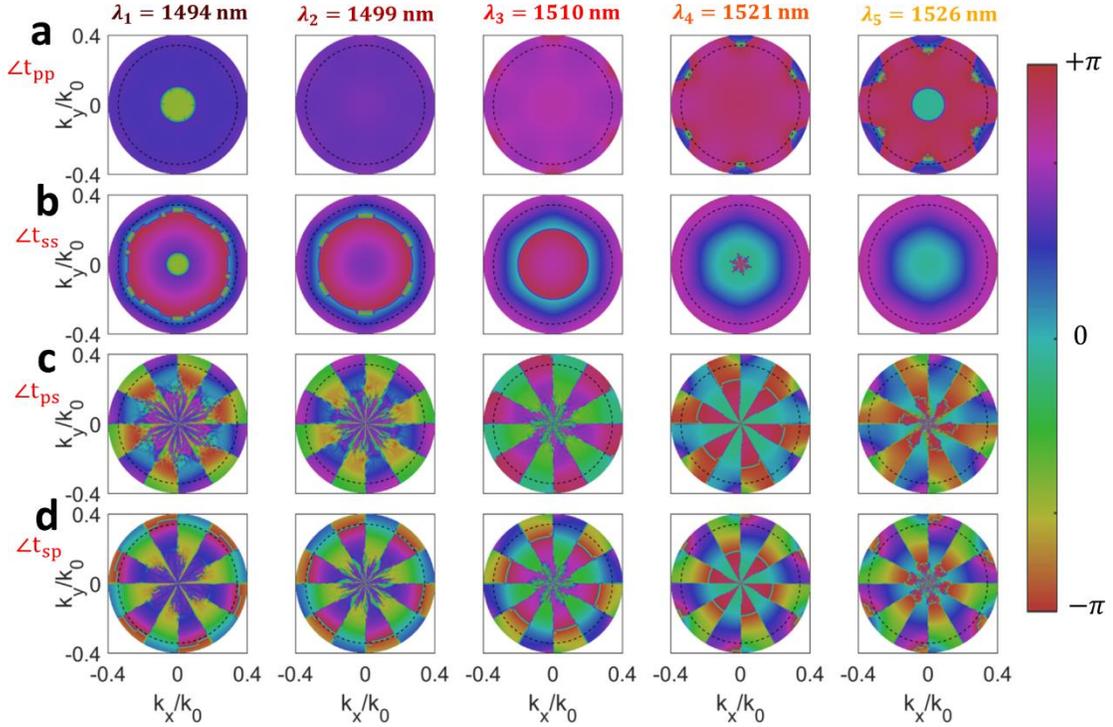

**Figure S3.** Calculated phases of the complex co- and cross-polarized transfer functions for the five wavelengths considered in Fig. 3 of the main text. Each column corresponds to a different wavelength, reported on top of the column. (a) Co-polarized transfer functions $\angle t_{pp}$. (b) Co-polarized transfer functions $\angle t_{ss}$. (c) Cross-polarized transfer functions $\angle t_{ps}$. (d) Cross-polarized transfer functions $\angle t_{sp}$.

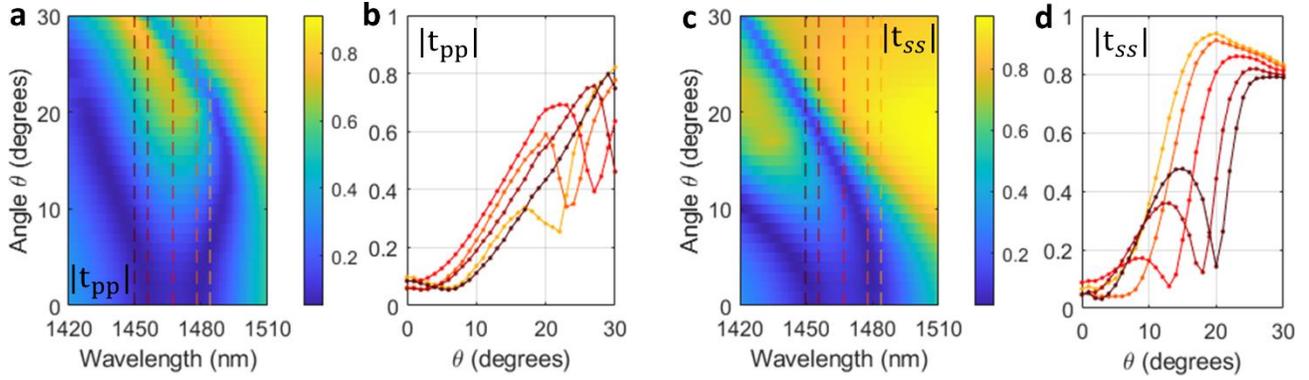

**Figure S4.** (a) Measured p-polarized transmission amplitude versus the polar angle $\theta$ and impinging wavelength, and for azimuthal angle $\phi = 30^o$. (b) Vertical cuts from panel a, corresponding to the color-coded dashed vertical lines. (c-d) Same as in (a-b), but for the s-polarized transmission amplitude.

### S2.2 Measured transmission amplitudes for $\phi = 30^o$

Figures 2(g-j) of the main text show the measured p- and s-polarized transmission amplitudes versus wavelength and polar angle $\theta$, for a fixed value of the azimuthal angle $\phi = 0^o$. For completeness, in Fig. S4 we show additional measured data taken at $\phi = 30^o$. These data agree very well with the corresponding simulated transfer functions (not shown here).

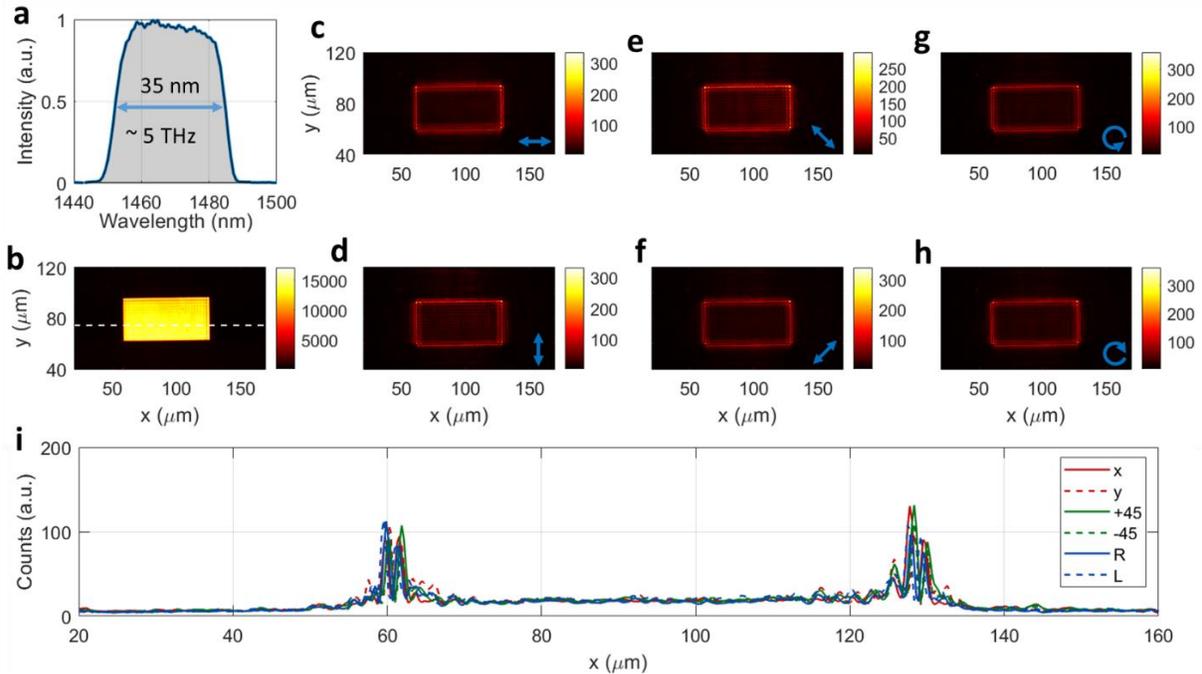

**Figure S5.** Edge detection with a rectangular target. (a) Spectrum of the excitation. (b) Unfiltered image. (c-h) Output images when the metasurface is placed in front of the target and for six different polarizations of the input light: linearly polarized along (c) x, (d) y, (e) the x-y diagonal, (f) the x-y anti-diagonal, or (g) right and (h) left circularly polarized. (i) Horizontal cuts of the plots in panels (c-h) (as indicated in the legend), corresponding to the vertical position denoted by the white dashed line in panel b.

## S2.3 Edge detection with rectangular targets

In Fig. 5 of the main text we demonstrated broadband and polarization-independent edge detection using a CUNY logo as a test image. In Fig. S5 we show additional imaging measurements done with a rectangular shape with dimensions 70x35 µm. All experimental conditions, including the excitation spectrum, are the same as in Fig. 5 of the main text.

## S3. Image processing with metasurfaces – theoretical calculations

In this paragraph we outline the mathematical steps to calculate the image processing imparted by a generic metasurface. We assume (Fig. S6) that an optical image is defined in the plane z = 0 by an intensity profile $I_{in}(x, y) = |\mathbf{E}_{in}(x, y)|^2$, where $\mathbf{E}_{in}(x, y) = E_{in}(x, y)\mathbf{e}_{in}$ is an electric field with polarization direction $\mathbf{e}_{in}$ and angular frequency $\omega = 2\pi c/\lambda = k_0 c$. For concreteness, the image can be thought as being generated by a plane wave with polarization $\mathbf{e}_{in}$ impinging on an aperture, but we notice that the calculations shown here, and the general concept of image processing, are independent of the way in which the image is created. Following standard Fourier optics [1], the image can be decomposed into a bundle of plane waves, each propagating along a direction identified by the polar and azimuthal angles $\theta$ and $\phi$. In particular, assuming that the plane wave impinging on the aperture in Fig. S6 has electric field $\mathbf{E}_0 = [E_{0,x}, E_{0,y}, 0]^T$, the field generated at a point identified by the spherical coordinates $(r, \theta, \phi)$ is given by [2]

$$\mathbf{E}(r,\theta,\phi) = ik_0 \frac{e^{-ik_0 r}}{2\pi r} f_{in}(k_x, k_y)[\mathbf{e}_\theta (E_{0,x} \cos\phi + E_{0,y} \sin\phi) + \mathbf{e}_\phi \cos\theta (E_{0,y} \cos\phi - E_{0,x} \sin\phi)] \quad (S1)$$

where $f_{in}(k_x, k_y) \equiv \int dx dy e^{-i(k_x x + k_y y)} E_{in}(x, y)$ is the Fourier transform of the input image. Thus, in the far field of the image ($r \gg \lambda$), the field propagating along each direction $(\theta, \phi)$ is given (up to an overall constant factor) by the plane wave

$$\mathbf{E}_{in}(\theta, \phi) = f_{in}(k_x, k_y)[\mathbf{e}_p E_p(\theta, \phi) + \mathbf{e}_s E_s(\theta, \phi)] \quad (S2)$$

where we identified the directional vectors of s and p polarization, $\mathbf{e}_p = \mathbf{e}_\theta$ and $\mathbf{e}_s = \mathbf{e}_\phi$, and defined $E_p(\theta,\phi) \equiv E_{0,x} \cos\phi + E_{0,y} \sin\phi$ and $E_s(\theta,\phi) \equiv \cos\theta (E_{0,y} \cos\phi - E_{0,x} \sin\phi)$. In Eqs. (S1)-(S2) and all equations below it is always assumed that the wave vector components $[k_x, k_y]$ depend on the angles $(\theta, \phi)$ through the standard coordinate transfomation $[k_x, k_y] = k_0 \sin\theta[\cos\phi, \sin\phi]$. The polarization of each wave in Eq. (S2) is generally a mixture of s and p polarization, depending on the polarization of the illumination and the direction of propagation. The response of the metasurface can be described by a 2x2 matrix of transfer functions, describing the angle-dependent co-polarized and cross-polarized transmission coefficients

$$t(\theta,\phi) = \begin{pmatrix} t_{ss}(\theta,\phi) & t_{sp}(\theta,\phi) \\ t_{ps}(\theta,\phi) & t_{pp}(\theta,\phi) \end{pmatrix}. \tag{S3}$$

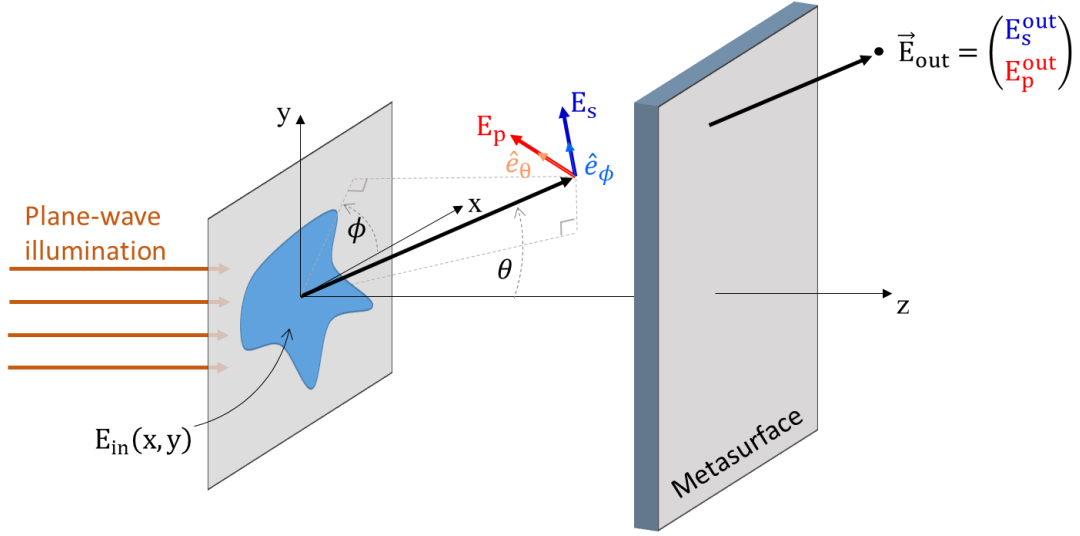

**Figure S6.** Schematic of the fields scattered by the image and filtered by the metasurface. See text for details.

Specifically, the field transmitted through the metasurface at any given angle $(\theta,\phi)$ is

$$\mathbf{E}_{out}(\theta,\phi) = \begin{pmatrix} E_s^{out}(\theta,\phi) \\ E_p^{out}(\theta,\phi) \end{pmatrix} = f_{in}(k_x,k_y) \begin{pmatrix} t_{ss}(\theta,\phi) & t_{sp}(\theta,\phi) \\ t_{ps}(\theta,\phi) & t_{pp}(\theta,\phi) \end{pmatrix} \begin{pmatrix} E_s(\theta,\phi) \\ E_p(\theta,\phi) \end{pmatrix} \tag{S4}$$

In order to calculate the image generated by this filtered bundle of waves, we project them back to the z = 0 plane. This is equivalent to collecting and re-focusing these waves on a plane placed at a z = 4f with a pair of identical lenses with focal length f. Moreover, we transform the field into the x-y polarization basis. The overall transformation corresponds to the inverse of the plane-wave expansion in Eq. (S1). That is, apart from an overall proportionality factor,

$$\begin{pmatrix} E_x^{out}(\theta,\phi) \\ E_y^{out}(\theta,\phi) \end{pmatrix} = \overline{\overline{M}}^{-1}(\theta,\phi) \begin{pmatrix} E_s^{out}(\theta,\phi) \\ E_p^{out}(\theta,\phi) \end{pmatrix} \tag{S5}$$

where $\overline{\overline{M}}^{-1}$ is the inverse of the matrix

$$\overline{\overline{M}}(\theta,\phi) = \begin{pmatrix} \cos\phi & \sin\phi \\ -\cos\theta\sin\phi & \cos\theta\cos\phi \end{pmatrix}. \tag{S6}$$

Finally, the spatially dependent output fields $E_x^{out}(x,y)$ and $E_y^{out}(x,y)$, corresponding to the electric field of the filtered image, are obtained via the inverse Fourier transform

$$E_{x/y}^{out}(x,y) = \frac{1}{2\pi}\int dxdy\, e^{i(k_x x+k_y y)} E_{x/y}^{out}(k_x,k_y), \tag{S7}$$

and the intensity profile is then calculated via $I(x,y) = |E_x^{out}(x,y)|^2 + |E_y^{out}(x,y)|^2$. The calculations shown in Fig. 3 of the main text have been performed assuming an unpolarized excitation. To emulate the unpolarized excitation we repeated the calculations outlined above twice, assuming first an x-polarized excitation ($E_{0,x}=1, E_{0,y}=0$) and then a y-polarized excitation ($E_{0,x}=0, E_{0,y}=1$). We then averaged the filtered intensity maps $I(x,y)$ obtained in the two excitation scenarios.

Equations S4-S7 can be used to readily show that, when Eq. 1 in the main text is valid, polarization-independent second order differentiation is achieved. When Eq. 1 is valid,

$$t(\theta,\phi) = \begin{pmatrix} t_{ss}(\theta,\phi) & t_{sp}(\theta,\phi) \\ t_{ps}(\theta,\phi) & t_{pp}(\theta,\phi) \end{pmatrix} = \begin{pmatrix} C\sin^2\theta & 0 \\ 0 & C\sin^2\theta \end{pmatrix}, \tag{S8}$$

the output fields in Eq. S4 become (apart from an overall constant)

$$\mathbf{E}_{out}(\theta,\phi) = \begin{pmatrix} E_s^{out}(\theta,\phi) \\ E_p^{out}(\theta,\phi) \end{pmatrix} = \sin^2\theta \cdot f_{in}(k_x,k_y) \cdot \overline{\overline{M}}(\theta,\phi)\begin{pmatrix} E_{0,x} \\ E_{0,y} \end{pmatrix}. \tag{S9}$$

Thus, the filtered fields in the xy cartesian basis (Eq. S5) become

$$\begin{pmatrix} E_x^{out}(\theta,\phi) \\ E_y^{out}(\theta,\phi) \end{pmatrix} = \sin^2\theta \cdot f_{in}(k_x,k_y)\begin{pmatrix} E_{0,x} \\ E_{0,y} \end{pmatrix}. \tag{S10}$$

Finally, by applying Eq. S7 we find

$$\begin{pmatrix} E_x^{out}(x,y) \\ E_y^{out}(x,y) \end{pmatrix} \propto \nabla^2 E_{in}(x,y)\begin{pmatrix} E_{0,x} \\ E_{0,y} \end{pmatrix}. \tag{S11}$$

Equation S11 shows that any polarization of the input image (either linear, circular, elliptical or arbitrary mixture of them) will be equally processed by the metasurface, and the total spatial intensity map $I(x,y) = |E_x^{out}(x,y)|^2 + |E_y^{out}(x,y)|^2$ will be independent of the input polarization.

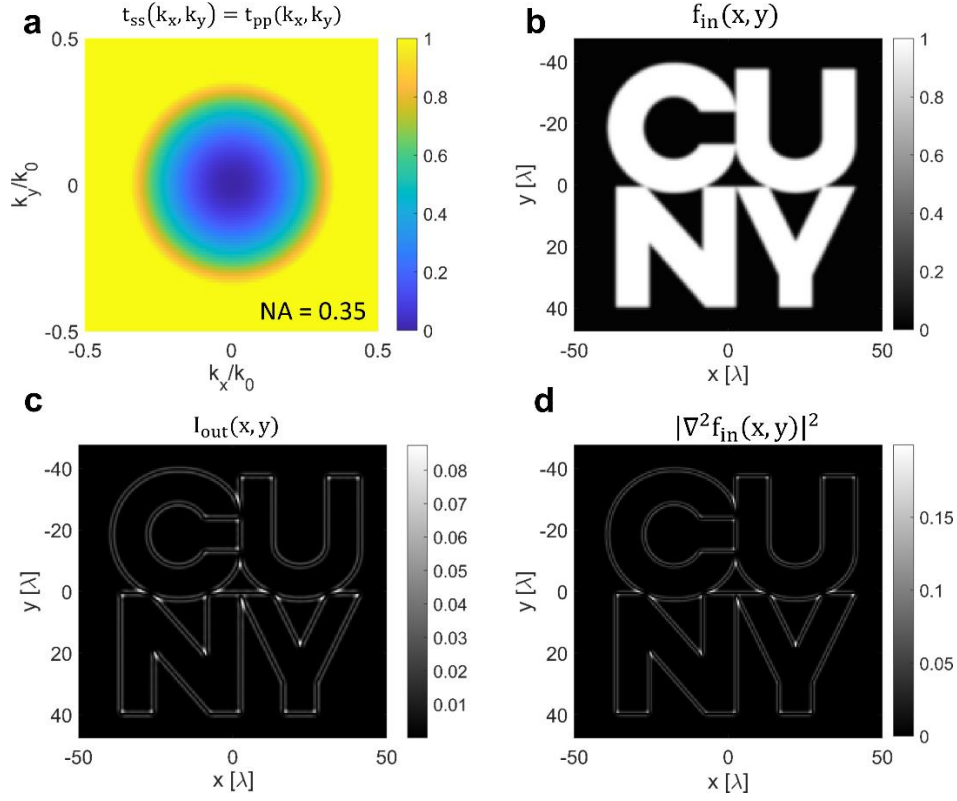

**Figure S7. Calculated image processing assuming an ideal polarization-independent filter.** (a) Co-polarized ideal transfer functions $t_{ss}(k_x, k_y) = t_{pp}(k_x, k_y)$ of the ideal filter (see text for details. (b) Input image. The shape and dimensions are almost identical to the target used in the experiments in Figs. 4 and 5 of the main text. (c) Output image calculated assuming the transfer function in panel a. (d) Numerically calculated Laplacian of the input image.

## S4. Maximum achievable efficiency

In our experiments the peak efficiency $\eta_{peak}$, defined as the ratio between the peak intensities in the output and input images, is about 5%-10% in the monochromatic excitation case (Fig. 4 of the main text), and about 3.5% in the broadband excitation scenario (Fig. 5 of the main text). While these numbers might seem small, here we show that they are actually very close to the maximum efficiency obtainable for an ideal polarization-independent k-space filter performing edge detection at a fixed NA. To demonstrate this, we consider an ideal filter described by identical co-polarized transfer functions $t_{ss}(k_x, k_y) = t_{pp}(k_x, k_y) = t_{ideal}(k_x, k_y)$ and zero cross-polarized transfer functions, $t_{sp}(k_x, k_y) = t_{ps}(k_x, k_y) = 0$. We assume that the ideal transfer function is given by

$$t_{ideal}(k_x, k_y) = \begin{cases} \left(\frac{1}{k_0^2 NA^2}\right) \cdot (k_x^2 + k_y^2) & \text{if } k_x^2 + k_y^2 \leq k_0^2 NA^2 \\ 1 & \text{otherwise} \end{cases}.$$

This transfer function provides the required Laplacian response up to a spatial frequency corresponding to $\sqrt{k_x^2 + k_y^2} = k_0 NA$, where the transfer function reaches 1. The transmission then remains one for wavevectors with moduli larger than $k_0 NA$. This device acts as an ideal edge-detectors for images whose smallest spatial features correspond to wavevectors with moduli equal to $k_0 NA$. In Fig, S7a we consider this ideal transfer function for the case of NA = 0.35, i.e. the same NA achieved by our device experimentally. We then consider an input image (Fig. S7b) with the same shape and very similar dimensions as the one considered in the experiment, and normalized such that the maximum intensity is 1. Following the formalism described in section S4, we calculate the expected output image (Figs. S7c) assuming that the input image is processed by an ideal filter with the transfer functions shown in Fig. S7a. As expected, a clear edge enhancement occurs in the output image (Figs. S7c). The maximum intensity in the output image is about 8%-9% of the input intensity. These values matches well the typical values of 5%-10% obtained experimentally for monochromatic excitation. This shows that, for a given NA, the efficiency of our device is essentially the same as the one of an ideal edge-detector. This is also clear by comparing the output image in Fig. S7c with the image obtained by applying the exact Laplacian operator $\nabla^2 = \partial_x^2 + \partial_y^2$ to the same input image, which is shown in Fig. S7d. The peak intensity of $|\nabla^2 f_{in}(x,y)|^2$ is about 20% of the input intensity. Thus, the experimentally measured peak efficiencies $\eta_{peak} = 5\% - 10\%$ of our device are also close to the upper bound dictated by the intrinsic properties of the desired mathematical operation.

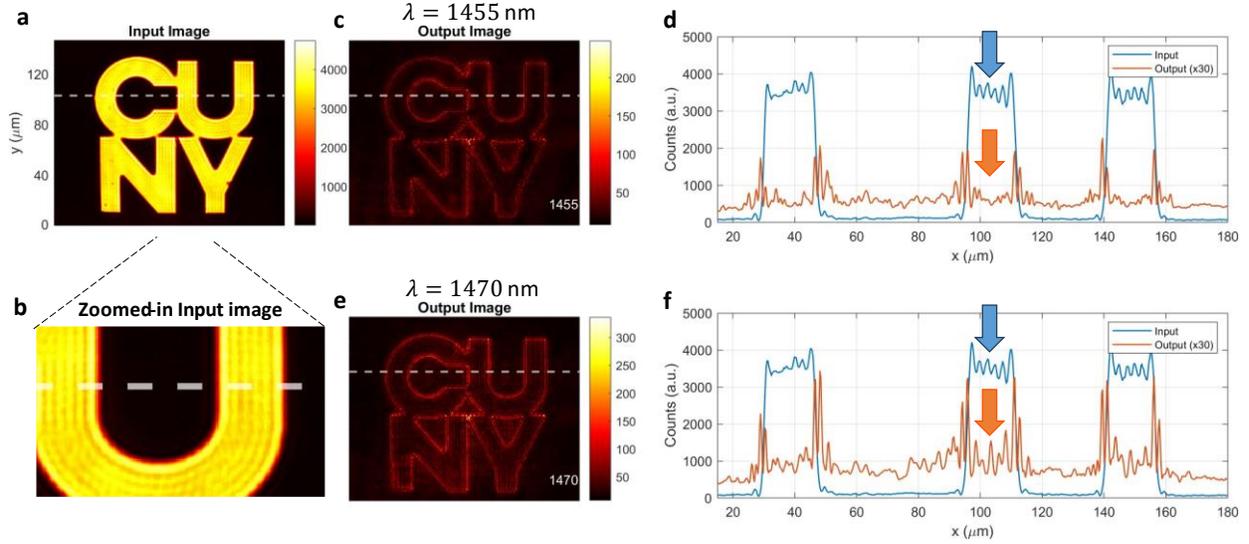

**Figure S8.** (a) Input image used in the experiment in Fig. 4 of the main text. (b) Zoomed-in portion of panel a, showing the weak intensity fluctuations within the bright areas. (c) Filtered image for an input wavelength of 1455 nm. (d) Horizontal cross-sectional cut of the input image (blue lines) and output image (orange lines), corresponding to the horizontal dashed lines in panels a and b. (e-f) Same as in panels (c-d) but for a wavelength of 1470 nm.

## S5. Origin of additional peaks in the output images

Figure 4 and 5 of the main text show the results of the edge-detection experiments. In all experimental figures, high-intensity peaks occur at the position of the main edges of the figure, surrounded by a much weaker background. In particular, some weaker peaks are visible in the experimental plots in Fig. 4c and 5i Several effects contribute to the creation of these additional peaks and to the background, as described in this section.

While some portion of the background is due to unavoidable noise (e.g. in the camera), most of the additional peaks have actually a real physical origin, rooted in the fact that edge-detection is implemented here via the mathematical operation of Laplacian differentiation, i.e., second order derivative.

Our input images contain some "strong edges", i.e., spatial regions where the optical intensity suddenly varies from low to high values, which lead to large absolute values of the second order derivatives and thus to large-intensity peaks in the output images. In particular, each strong edge in the input image will result in two peaks in the output image, as expected from the second order differentiation of a step-like function. In a practical scenario, the presence of two peaks (instead of one) for each edge does not introduce any detrimental effect – in fact, it can be used to find the exact spatial position of each edge even more accurately.

Moreover, in our input images the intensity is not perfectly flat within the bright areas. In Fig. S8 we reproduced some of the panels of Fig. 4, together with additional cross-sectional cuts. Weak spatial fluctuations of the intensity are clearly visible both in the 2D plot in Figs. S8(a-b), and in the 1D cuts (blue lines) in Figs. S8d and S8f, denoted by the blue arrows. These spatial intensity fluctuations are due to diffraction of light at the metallic apertures that are used in our experiment to generate the input image. Since our metasurface performs a mathematical differentiation on the whole input image, these weaker spatial variations will be differentiated as well, resulting in a set of weaker peaks in the output image. Some examples of these peaks are indicated by the orange arrows in Fig. S8(d,f).

From a practical point of view, the presence of these additional weaker peaks does not limit the capability of detecting the position of the "strong edges", since the intensity of each peak in the output image is always proportional to the spatial derivative (at the same position) of the input image. Thus, the sharpest edges will always correspond to the two strongest peaks. In fact, the fact that we can experimentally detect such weak intensity fluctuations is a further demonstration of the large quality and efficiency of our metasurface.

Besides the effect discuss above, other unwanted noise and additional signals (which does not necessarily have a peak-like structure) are due to a combination of the noise in the camera used for the experiment, and of the fact that the normal-incidence transmission of the metasurface, while remaining smaller than 1%, is not exactly zero. The latter effect implies that a very small portion of the 'DC component' of the input image is transferred into the output image without any differentiation. However, as clear from the experimental data in Figs. S8(d,f) and in the main text, this effect is very small, and it does not introduce any practical issue in the edge detection.

## S6. Increasing the bandwidth

In the main text we have demonstrated numerically and experimentally an edge-detecting metasurface with a bandwidth of about 5 THz around a central frequency of 198 THz. In this section, we show an example of another design, with further optimized dispersion engineering, which features a much larger bandwidth. The design has the same lattice constant as the one in the main text ($a$ = 924 nm), while slightly different slab thickness $H$ = 230 nm and hole radius $R$ = 280 nm. Fig. S9a shows the normal-incidence transmission spectrum of the device (orange line), compared to the spectrum of the device considered in the main text (blue line). The two optical modes, identified by the two frequencies at which the transmission is zero, are detuned by $\Delta \approx$ 7.5 THz. This, combined with increased optical linewidths, creates a range of B $\approx$ 10 THz where the transmission remains below 3%, ideally suited to suppress the DC component of the input image. The edge-detecting capability of this device is confirmed by the color-coded plot in Fig. S9b, which shows the p-polarized transmission versus polar angle theta at several wavelengths across the bandwidth. At all frequencies within a B ≈ 10 THz centered around 210 THz (68 nm at 1427 nm) the transmission is low at small angles, and it progressively increases as the angle $\theta$ increases, albeit with slightly different NAs.

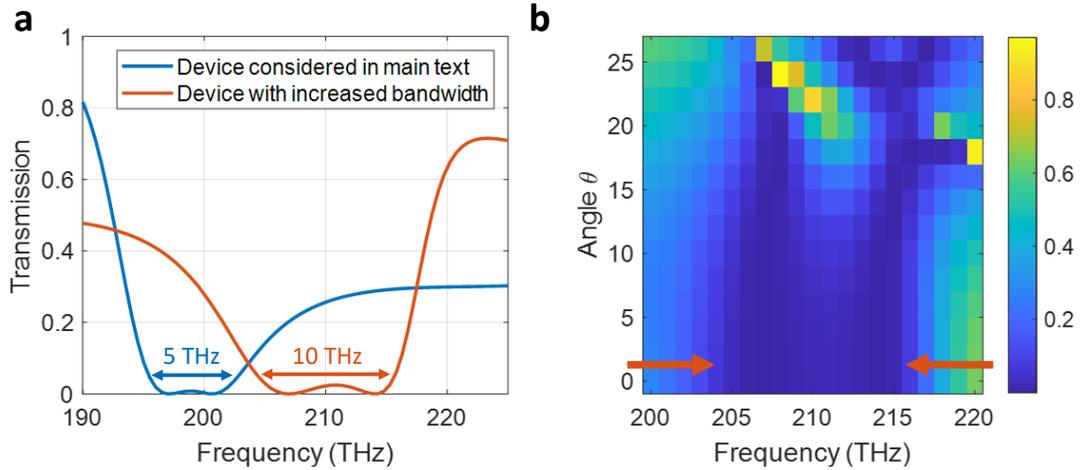

**Figure S9.** (a) Normal-incidence transmission spectrum of the device considered in the main text (blue line) and of a slightly different design (orange line), where the optical modes are further detuned from each other, resulting in band of about 10 THz where the transmission remains below 3%. (b) p-polarized transmission of this new design as a function of the angle $\theta$ and impinging frequency.

## Supplementary References